\documentclass[journal,comsoc,twoside]{IEEEtran}

\usepackage{amssymb}

\usepackage{graphicx,cite,epsfig,amssymb,amsmath}
\usepackage{epstopdf}

\usepackage{pifont}
\usepackage{stmaryrd}
\usepackage{bbding}
\usepackage{algorithm}
\usepackage{algorithmic}
\usepackage{mathrsfs}
\usepackage{latexsym}
\usepackage{indentfirst}
\usepackage{fmtcount}
\usepackage{cite}
\usepackage{float}
\usepackage{morefloats}

\usepackage{caption}
\usepackage{subcaption}

\usepackage{color}
\usepackage{mathtools}
\usepackage{xfrac}

\usepackage{array,colortbl,multirow,multicol,booktabs,ctable,bigstrut}

\usepackage[
top    = 0.75in,
bottom = 0.75in,
left   = 0.75in,
right  = 0.75in]{geometry}

\begin{document}
\title{AI Coding: Learning to Construct Error Correction Codes}

\author{Lingchen~Huang,~\IEEEmembership{Member,~IEEE,}
        Huazi~Zhang,~\IEEEmembership{Member,~IEEE,}
        Rong~Li,~\IEEEmembership{Member,~IEEE,}
        Yiqun~Ge,~\IEEEmembership{Member,~IEEE,}
        Jun~Wang,~\IEEEmembership{Member,~IEEE,}
    \thanks{L.~Huang, H.~Zhang, R.~Li and J.~Wang are with the Hangzhou Research Center, Huawei Technologies, Hangzhou, China.
    Y.~Ge is with the Ottawa Research Center, Huawei Technologies, Ottawa, Canada.
    (Email: \{huanglingchen, zhanghuazi, lirongone.li, yiqun.ge, justin.wangjun\}@huawei.com).}
}

\markboth{IEEE TRANSACTIONS ON COMMUNICATIONS, VOL.~XX, NO.~X, OCTOBER~2019}{Huang \MakeLowercase{\textit{et al.}}: AI Coding: Learning to Construct Error Correction Codes}

\maketitle

\begin{abstract}
In this paper, we investigate an artificial-intelligence (AI) driven approach to design error correction codes (ECC). Classic error-correction code design based upon coding-theoretic principles typically strives to optimize some performance-related code property such as minimum Hamming distance, decoding threshold, or subchannel reliability ordering. In contrast, AI-driven approaches, such as reinforcement learning (RL) and genetic algorithms, rely primarily on optimization methods to learn the parameters of an optimal code within a certain code family. We employ a constructor-evaluator framework, in which the code constructor can be realized by various AI algorithms and the code evaluator provides code performance metric measurements. The code constructor keeps improving the code construction to maximize code performance that is evaluated by the code evaluator. As examples, we focus on RL and genetic algorithms to construct linear block codes and polar codes. The results show that comparable code performance can be achieved with respect to the existing codes. It is noteworthy that our method can provide superior performances to classic constructions in certain cases (e.g., list decoding for polar codes).
\end{abstract}


\IEEEpeerreviewmaketitle

\section{Introduction}\label{section:intro}
Error correction codes (ECC) have been widely used in communication systems for the data transmission over unreliable or noisy channels.
In~\cite{Shannon}, Shannon provided the definition of channel capacity and proved the channel coding theorem,
\begin{quotation}
``\emph{All rates below capacity $C$ are achievable, that is, for arbitrary small $\epsilon>0$ and rate $R<C$, there exists a coding system with maximum probability of error
  \begin{equation}\label{equ:coding_theorem}
    \lambda \leq \epsilon
  \end{equation}
for sufficiently large code length $n$. Conversely, if $\lambda \rightarrow 0$, then $R\leq C$.''}
\end{quotation}
Following Shannon's work, great effort has been continuously devoted to designing ECC and their decoding algorithms to achieve or approach the channel capacity.

Specifically, with a code rate smaller than channel capacity, code construction and decoding algorithms are designed to improve its code performance. Equivalently, given a target error rate, we optimize code design to maximize the achievable code rate, i.e. to approach the channel capacity.

\begin{figure}
\centering
    \includegraphics[width = 0.4\textwidth]{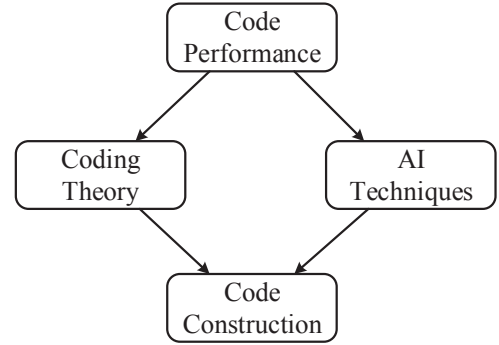}
    \caption{Error correction code design logic}
    \label{fig:construction_logic_classic}
\end{figure}

\subsection{Code design based on coding theory}
Classical code construction design is built upon coding theory, in which code performance is analytically derived in terms of various types of code properties.
To tune these properties is to control the code performance so that code design problems are translated into code property optimization problems.

\emph{Hamming distance} is an important code property for linear block codes of all lengths.
For short codes, it is the dominant factor in performance, when maximum-likelihood (ML) decoding is feasible.
For long codes, it is also important for performance in the high signal-to-noise ratio (SNR) regime.
A linear block code can be defined by a generator matrix $\bf{G}$ or the corresponding parity check matrix $\bf{H}$ over finite fields.
Directed by the knowledge of finite field algebra, the distance profile of linear block codes can be optimized, and in particular, the minimum distance.
Examples include Hamming codes, Golay codes, Reed-Muller (RM) codes, quadratic residue (QR) codes, Bose-Chaudhuri-Hocquenghem (BCH) codes, Reed-Solomon (RS) codes, etc.

Similar to the Hamming distance profile, \emph{free distance}, another code property, is targeted for convolutional codes.
Convolutional codes~\cite{CC:Elias} are characterized by code rate and the memory order of the encoder.
By increasing the memory order and selecting proper polynomials, larger free distance can be obtained at the expense of encoding and decoding complexity.
Turbo codes~\cite{Turbo}, by concatenating convolutional codes in parallel, are the first capacity-approaching codes under iterative decoding.

In addition to the distance profile, code properties of \emph{decoding threshold} and \emph{girth} are adopted for the design of low density parity check (LDPC) codes.
First investigated in 1962~\cite{LDPC:Gallager}, they are defined by low density parity check matrices, or equivalently, Tanner graphs.
Three decades later, LDPC codes were re-discovered~\cite{LDPC:MacKay} and shown to approach the capacity with belief propagation (BP) decoding on sparse Tanner graphs.
Code structures, such as cyclic and quasi-cyclic (QC)~\cite{LDPC:QC:Fossorier}, not only provide minimum distance guarantee but also simplify hardware implementation.
The most relevant code property, however, is the decoding threshold~\cite{LDPC:Richardson}.
Assuming BP decoding on a cycle-free Tanner graph, it can accurately predict the asymptotic performance.
The decoding threshold can be obtained by the extrinsic information transfer (EXIT) chart~\cite{LDPC:tenBrink} technique, as well as density evolution (DE)~\cite{LDPC:Richardson,LDPC:RichardsonBook} and its Gaussian approximation (GA)~\cite{LDPC:Chung}, as a function of the check node and variable node degree distributions of the parity check matrix.
Thus, the degree distributions of a code ensemble can be optimized.
In addition, the girth, defined as the minimum cycle length in the Tanner graph, is maximized to maximally satisfy the cycle-free assumption.

The code property of \emph{synthesized subchannel reliabilities} can be targeted for the design of polar codes~\cite{Polar:Arikan}, the first class of capacity-achieving codes with successive cancellation (SC) decoding.
For polar codes, physical channels are synthesized to $N$ polarized subchannels, with the $K$ most reliable ones selected to carry information bits.
As $N$ increases, subchannels polarize to either purely noiseless or completely noisy, where the fraction of noiseless subchannels approaches the channel capacity~\cite{Polar:Arikan}.
For binary erasure channel (BEC), subchannel reliabilities can be efficiently calculated by Bhattacharyya parameters.
For general binary-input memoryless channels, DE was applied to calculate subchannel reliabilities~\cite{Polar:DE1_Mori,Polar:DE2_Mori}, and then improved in~\cite{Polar:DE3_Tal} and analyzed in~\cite{Polar:DE4_Pedarsani} in terms of complexity.
For AWGN channels, GA was proposed~\cite{Polar:GA_Trifonov} to further reduce complexity with negligible performance loss.
Recently, a polarization weight (PW) method~\cite{Polar:PW,Polar:betaExpansion} was proposed to generate a universal reliability order for all code rates, lengths and channel conditions. Such a channel-independent design principle is adopted by 5G in the form of a length-1024 reliability sequence~\cite{Polar:HW_seq}.

As concluded in Fig.~\ref{fig:construction_logic_classic} (left branch), the classical code design philosophy relies on coding theory (e.g., finite field theory, information theory) as a bridge between code performance and code construction.

\subsection{Code design based on AI}\label{section:intro:AI}
Recently, AI techniques have been widely applied to many industry and research domains, thanks to advances in algorithms, an abundance of data, and improvements in computational capabilities.

In communication systems, AI-driven transceivers have been studied.
By treating an end-to-end communication system as an autoencoder, people proposed to optimize an entire transceiver jointly given a channel model without any expert knowledge about channel coding and modulation~\cite{AI:AE}.
In addition, for AWGN channel with feedback, recurrent neural network (RNN) was used to jointly optimize the encoding and decoding~\cite{AI:DeepCode}.
By regarding a channel decoder as a classifier, it was reported that a one-shot NN-based decoding could approach maximum a posteriori (MAP) performance for short codes~\cite{AI:decoding}.
It was also observed that for structured codes, the NN-based decoder can be generalized to some untrained codewords, even though no knowledge about the code structure is explicitly taught.
However, this NN-based decoder is a classifier in nature, and its complexity is proportional to the number of codewords to be classified. As the code space expands exponentially with the code length, the NN-based decoder may not satisfy the stringent latency constraint in physical layer.

In contrast, our work focuses on using AI techniques to help design codes rather than to directly encode and decode signals.
Code design can be done offline where the latency constraint is significantly relaxed.
Moreover, we can continue to use legacy encoding and decoding algorithms as they admit efficient and flexible hardware or software implementations.

As shown in Fig.~\ref{fig:construction_logic_classic} (right branch), we explore the alternative AI techniques for code construction, in addition to the expert knowledge from coding theory. There are numerous AI algorithms out there, which could not be all covered in this paper. We will focus on reinforcement learning (RL) and genetic algorithm as representatives. Nevertheless, we work within a general framework that may accommodate various AI algorithms.

Specifically, we hope to answer the following two questions:
\begin{itemize}
  \item Q1: Can AI algorithms independently learn a code construction (or part of it) within a given general encoding/decoding framework?
  \item Q2: Can the learned codes achieve comparable or better error correction performance with respect to those derived in classic coding theory?
\end{itemize}

For the first question, we try to utilize AI techniques to learn linear code constructions.
By code construction we mean the degree of freedom we have in determining a set of codes beyond the minimum constraints required to specify a general scope of codes.
The input to AI is restricted to the code performance metric measured by an evaluator (viewed as a black box) that implements off-the-shelf decoders under a specific channel condition.
Therefore, AI knows neither the internal mechanism of the decoders nor the code properties so that code construction beyond the encoding/decoding constraints is learned without the expert knowledge in coding theory.

For the second question, we demonstrate that AI-driven techniques provide solutions that perform better when expert knowledge fails to guarantee optimal code performance. These cases include (i) a target decoder is impractical, or (ii) the relation between code performance and code properties is not theoretically analyzable (incomplete or inaccurate).
For example, although the minimum distance of short linear block codes can be optimized, the resulting optimality of code performance is only guaranteed under maximum likelihood (ML) decoding at high SNRs. In reality, ML decoders may be impractical to implement, and a high SNR may not always be available.
As for polar codes, existing theoretical analysis mainly focuses on SC decoders.
To the best of our knowledge, the widely deployed successive cancellation list (SCL) decoders still lack a rigorous performance analysis, even though there have been some heuristic constructions optimized for SCL~\cite{Polar:RM_Polar_Li,Polar:BP_LLR_Qin}, and its variants CRC-aided SCL (CA-SCL)~\cite{Polar:HW_seq} and parity-check SCL~\cite{Polar:Subcode_Trifonov,Polar:PCC_Wang,Polar:PC5G_Huawei}.
In the absence of theoretical guarantee, the open question is whether the current code constructions are optimal (for a specific decoder)?
We will address these cases and show that AI techniques can deliver comparable or better performances.

It is worth noting that a recent approach \cite{Polar:GeneAlg} also focuses on using AI techniques for code design rather than decoding. In their work, the off-the-shelf polar decoder is embedded in the code optimization loop. Only the code construction was optimized while the encoding and decoding methods remain the same, which allows efficient implementation of legacy encoding and decoding algorithms. Specifically, the proposed code optimization method is based on the genetic algorithm, where code constructions evolve via evolutionary transformations based on their error performance.

In this paper, we employ a general constructor-evaluator framework to design error correction codes.
AI techniques are investigated under this framework. The performance of constructed codes are compared with the state-of-the-art classical ones, and the results are discussed.
The structure of this paper is as follows.
Section~\ref{section:learning} introduces the constructor-evaluator framework. By instance, the constructor is implemented by RL and genetic algorithms. The evaluator implements the decoder and channel conditions of interest and provides performance metric measurements for given code constructions.
Section~\ref{section:examples} shows examples of designing linear block codes and polar codes under the proposed framework.
Section~\ref{section:conclusion} concludes the paper and discusses some future works.

\section{Code construction based on learning}\label{section:learning}

\begin{figure}
\centering
    \includegraphics[width = 0.5\textwidth]{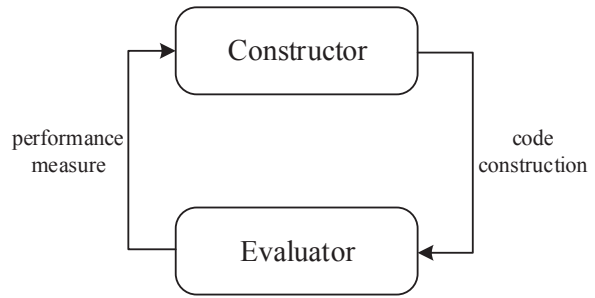}
    \caption{Constructor-evaluator framework}
    \label{fig:framework}
\end{figure}

\subsection{The constructor-evaluator framework}
We advocate a code design framework, as shown in Fig.~\ref{fig:framework}.
The framework consists of two parts, i.e. a code constructor and a code evaluator.
The code constructor iteratively learns a series of valid code constructions based on the performance metric feedback from the code evaluator.
The code evaluator provides an accurate performance metric calculation/estimation for a code construction under the decoder and channel condition of interest.
The code construction keeps improving through the iterative interactions between the constructor and evaluator until the performance metric converges.

The constructor-evaluator framework in Fig.~\ref{fig:framework} is quite general. The code constructor knows neither the internal mechanism nor the channel condition adopted by the code evaluator but requests the code evaluator to feed back an accurate performance metric of its current code construction under the evaluator-defined environment, by which the exploration of possible code construction opens for a wide range of decoding algorithms and channel conditions. Similar ideas have been proposed in existing works. For example, the differential evolution algorithm was used to optimize the degree distribution of LDPC codes under both erasure channel \cite{DiffEvo:Shokrollahi} and AWGN channel \cite{LDPC:Richardson}. The algorithm also treats the optimization problem as a black box, and merely relies on the cost function (e.g., decoding threshold) as feedback.

In most cases, the code constructor is trained offline, because both coded bits and decoding results can be efficiently generated and training computation is not an issue in an offline simulator. Once constructed e.g., the performance metric converges, the resultant codes can be directly implemented in a low-complexity practical system with legacy encoding and decoding algorithms, which is applicable from the industry's point of view.

Note that the constructed codes are closely related to the code evaluator because of performance metric.
Different code evaluators, e.g. with different decoding parameters, channel condition assumptions, or decoding algorithms, may result in different code constructions in the end.
During training procedure, we choose some realistic decoding algorithms and channel condition.
In theory, an online training is a natural extension by collecting performance metric samples from a real-world decoder and continuing to improve code construction.

\subsection{Constructor}\label{section:learning:constructor}
The code constructor generates code constructions based on performance metrics feedback from the code evaluator.

To fit the code design problem into the AI algorithms, a code construction is defined under the constraints of code representations. For example,
\begin{itemize}
  \item \textbf{Binary matrix}: the generator matrix for any linear block codes, or the parity-check matrix for LDPC codes. This is the most general form of definition.
  \item \textbf{Binary vector}: a more succinct form of definition for some codes, including the generator polynomials for convolutional codes, or whether a synthesized subchannel is frozen for polar codes.
  \item \textbf{Nested representation}: defining a set of codes in a set of nested matrices or vectors. This bears practical importance due to low implementation complexity and rate compatibility. Examples include LTE-RM codes and 5G Polar codes.
\end{itemize}

According to the code representations and how the construction improvement procedure is modeled, there are several approaches to implement the constructor:

\subsubsection{Reinforcement learning approach}\label{section:learning:constructor:RL}
RL approach can be used, because we model construction procedure as a Markov decision process (MDP).
An MDP is defined by a 4-tuple ($S$, $A$, $P_{a}$, $R$):
\begin{itemize}
  \item $S$ is the state space,
  \item $A$ is the action space,
  \item $P_{a}(s,s')=Pr(s_{t+1}=s'|s_t=s,a_t=a)$ is the probability that action $a$ in state $s$ at time $t$ will lead to state $s'$ at time $t+1$,
  \item $R$ is the immediate reward feedback by code evaluator after transitioning from state $s$ to state $s'$, triggered by action $a$.
\end{itemize}

Code construction can be viewed as a decision process in general. For the binary matrix and binary vector code representations, the decisions correspond to which positions are 0 or 1. For the nested code representation, the decisions correspond to how to evolve from a set of subcodes to their supercodes (or vice versa).
In our setting, a state $s$ corresponds to a (valid) code construction, and an action $a$ that leads to the state transition $(s \rightarrow s')$ corresponds to a modification to the previous construction $s$.
The state transition $(s \rightarrow s')$ herein is deterministic by the action $a$ and the previous state $s$.
The reward function is the performance metric measurement with respect to the decoder and channel condition of interest.
In the end, a desired code construction can be obtained from the final state of the MDP.

There are several classical algorithms to solve MDP that are model-free, i.e., they do not require the model of the code evaluator. These include:
\begin{itemize}
  \item Q-learning~\cite{RL:Q-learning}: given the architecture of a code construction, its potential code construction schemes correspond to a finite set of discrete states and action spaces. In Q-learning, a table $Q(s,a)$ is maintained and updated to record an expected total reward metric to take action $a$ at state $s$.
    At learning stage, an $\epsilon$-greedy approach is often used to diverge the exploration in the state and action space.
    When a state transition  $(s,a,s',R)$ has been explored, the table $Q(s,a)$ can be updated by:
    \begin{equation}\label{equ:Q_learning}
    \Delta Q(s,a) = \alpha_{Q} \cdot [R+\gamma \cdot max_{a'}Q(s',a')-Q(s,a)],
    \end{equation}
    where $\alpha_{Q}$ is learning rate, $\gamma$ is reward discount factor, and $R$ is reward from the evaluator.
    After sufficient exploration, the table $Q(s,a)$ is then used to guide the MDP to maximize the total reward.
  \item Policy Gradient (PG)~\cite{RL:PG}: if the architecture of a code construction translates into an immense set of states and actions, we consider continuous state and action space. PG defines a differentiable policy function $\pi_{\theta_{PG}}(s,a)$, parameterized by $\theta_{PG}$, to select the action at each state.
    The policy function $\pi_{\theta_{PG}}(s,a)$ outputs a probability desity/mass function of taking each action $a$ at state $s$ according to the policy $\pi_{\theta_{PG}}$. Then the next state $s'$ is determined and the reward $R$ is evaluated by the code evaluator.
    When a complete episode $(s_0,a_0,R_0,s_1,a_1,R_1,\cdots,s_t,a_t,R_t,\cdots,s_{T-1},a_{T-1},R_{T-1})$ is explored, where $t$ is the time stamp and $T$ is the time horizon length, the policy function is updated:
    \begin{equation}\label{equ:PG}
    \Delta \theta_{PG} = \alpha_{PG} \cdot \sum_{t=0}^{T-1} [\nabla_{\theta_{PG}} \log \pi_{\theta_{PG}} (s_t,a_t) \cdot \sum_{t'=t}^{T-1} R_{t'}].
    \end{equation}
    After sufficient exploration, the policy function $\pi_{\theta_{PG}}$ can be used to lead the MDP to maximize the total reward.
  \item Advantage Actor Critic (A2C)~\cite{RL:A2C}: A2C merges the idea of state value function into PG to take advantage of stepwise update, and speeds up the convergence.
    In addition to the policy (actor) function $\pi_{\theta_A}(s,a)$, A2C defines a differentiable value (critic) function $V_{\theta_C}(s)$.
    The interaction between A2C and the code evaluator is similar to that of PG.
    For A2C, the policy (actor) update can be more frequent, i.e. in stepwise manner, since the cumulative reward from $s_t$, $\sum_{t'=t}^{T-1} R$, is estimated by the critic function.
    At each state translation exploration $(s,a,s',R)$, the advantage value is calculated:
    \begin{equation}\label{equ:A2C_Adv}
    Adv(s,s',R) = R+\gamma \cdot V_{\theta_C}(s') - V_{\theta_C}(s).
    \end{equation}
    Then the actor function $\pi_{\theta_A}$ can be adjusted by:
    \begin{equation}\label{equ:A2C_A}
    \Delta \theta_A = \alpha_A \cdot Adv(s,s',R) \cdot \nabla_{\theta_A} \log \pi_{\theta_A} (s,a).
    \end{equation}
    The critic function $V_{\theta_C}$ can be updated by:
    \begin{equation}\label{equ:A2C_C}
    \Delta \theta_C = \alpha_C \cdot Adv(s,s',R) \cdot \nabla_{\theta_C}V_{\theta_C}(s).
    \end{equation}
\end{itemize}

By viewing code construction as a decision process, its influence on code performance can be modeled (or approximated) as differentiable functions that can be realized by neural networks.
A code construction (to be optimized) is embedded in the coefficients of the neural networks. Due to the excellent function approximation capability of neural networks, these coefficients can be learned through optimizations techniques such as gradient descent.

\subsubsection{Genetic algorithm approach}\label{section:learning:constructor:genetic}
we observe that a code construction can usually be decomposed into many discrete decisions. For the binary matrix and binary vector code representations, the decisions on which positions are 0 or 1 can be made individually. These decisions may collaboratively contribute to the overall code performance. They resemble the ``chromosomes'' of a code construction, where a set of good decisions is likely to produce good code constructions. The process of refining these decisions can be defined in iterative steps, where each step produces a better construction based on several candidate constructions. Genetic algorithm is well-suited for this purpose.

Below, we briefly describe how a genetic algorithm can be applied to the code design.
\begin{enumerate}
  \item A number of code constructions $\{{\cal C}_{1},{\cal C}_{2},\cdots\}$ are randomly generated, defined as initial population.
  \item A subset of (good) code constructions are selected as parents, e.g., ${\cal C}_{p1}, {\cal C}_{p2}$.
  \item New constructions (offspring) are produced from crossover among parents, e.g., $\{{\cal C}_{p1}, {\cal C}_{p2}\} \rightarrow {\cal C}_{o1}$.
  \item Offsprings go through a random mutation to introduce new features, e.g., ${\cal C}_{o1} \rightarrow {\cal C}_{m1}$.
  \item Finally, good offspring replace bad ones in the population, and the process repeats.
\end{enumerate}

The above operations are defined in the context of error correction codes.
Regarding code definition ${\cal C}$, it may boil down to a set of chromosomes (binary vectors and matrices) accordingly.

Crossover is defined as taking part of the chromosomes from each parent, and combining them into an offspring.
This step resembles ``reproduction" in biological terms, in which offspring are expected to inherit some good properties from their parents.
Subsequently, mutation randomly alters some of the chromosomes to encourage exploration during evolution.

A fitness function is defined to indicate whether the newly produced offspring are good or bad.
In this work, the fitness is defined as the code performance.

\subsection{Evaluator}
The evaluator provides performance metric measurements for code constructions.
If the performance metric of the decoder is analyzable, it can be directly calculated.
In most cases, the error correction performance estimation in terms of block error rate (BLER) can be performed based on sampling techniques, such as Monte-Carlo (MC) method.
To ensure that the estimation is accurate and thereby does not mislead the constructor, sufficient MC sampling (simulation) should be performed to control the estimation confidence level. If the designed code is to work within a range of BLER level, the performance metric measurements can merge error correction performance at several SNR points. Measuring more than one SNR points provides a better control over the slope of BLER curve, at the cost of longer evaluation time. In addition, power consumption and implementation complexity for the encoding and the corresponding decoding also can be factored into the performance metric measurements.

Intuitively, the evaluator can be stationary. The decoding algorithm, including the parameters and the channel statistics can be static. Then the code design is preferred to be realized offline, and is not very sensitive to the code design time consumption. On the other hand, the evaluator can be non-stationary. For example, the channel statistics can be time-varying. Then online design may be required. In this case, a feedback link is required for performance measurement of each code construction. The communication cost and code design time consumption should be considered as well.

\section{Learning code representations}\label{section:examples}
In this section, we present several code design examples in which the code constructions are automatically generated by the constructor-evaluator framework. Specifically, we propose three types of AI algorithms to learn the code constructions under the three definitions mentioned in \ref{section:learning:constructor}:
\begin{itemize}
  \item \textbf{Binary matrix} $\rightarrow$ \textbf{policy gradient}: we provide an example of short linear block codes.
  \item \textbf{Binary vector} $\rightarrow$ \textbf{genetic algorithm}: we focus on polar codes with a fixed length and rate.
  \item \textbf{Nested representation} $\rightarrow$ \textbf{A2C algorithm}: we design nested polar codes within a range of code rates with the proposed scheme.
\end{itemize}
Although we provide three specific code examples, the AI algorithms are generic. That means all codes defined by the above three representations can be learned by the proposed methods.

We use the following hardware platform and software environment. For the reinforcement learning approach (including policy gradient and A2C), we use one Telsa V100 GPU and one 28-thread Xeon Gold CPU to accomplish learning within the Tensorflow framework.
For the genetic algorithm approach, the program is written in Matlab and C/C++ language and runs on a server that contains 4 Intel Xeon(R) E5-4627v2 (16M Cache, 3.30 GHz) CPUs with 12 cores and 256 GB RAM. For all the AI algorithms, we did not pay extra attention to the optimization of hyperparameters, as they seem not to affect the results very much.
For a proof of concept, binary codes and AWGN channel are considered in this work.
Codewords are encoded from randomly generated information bits.
It is shown that these learned codes can achieve at least as good error correction performance as the state-of-the-art codes.

\subsection{Binary matrix: linear block codes}
The definition of a linear block code of dimension $K$ and length $N$ is a binary generator matrix ${\bf G}$ in a standard form (i.e. ${\bf G}=[{\bf I},{\bf P}]$ where ${\bf I}$ is an identity matrix of size $K \times K$ and ${\bf P}$ is a matrix of size $K \times (N-K)$).
For the decoder, a class of most-reliable-basis (MRB) reprocessing based decoding can achieve near-ML performance.
The class of decoders uses the reliability measures at the decoder input to determine an information set consisting of the $K$ most reliable independent positions.
One efficient MRB reprocessing based decoder is the ordered statistic decoding (OSD)~\cite{MRB:OSD}.
By incorporating box and match algorithm (BMA) into OSD~\cite{MRB:BMA}, the computational complexity can be reduced at the cost of additional memory for a realistic implementation.
Therefore, our code evaluator deploys BMA decoder in this example.

\begin{figure}
\centering
    \includegraphics[width = 0.5\textwidth]{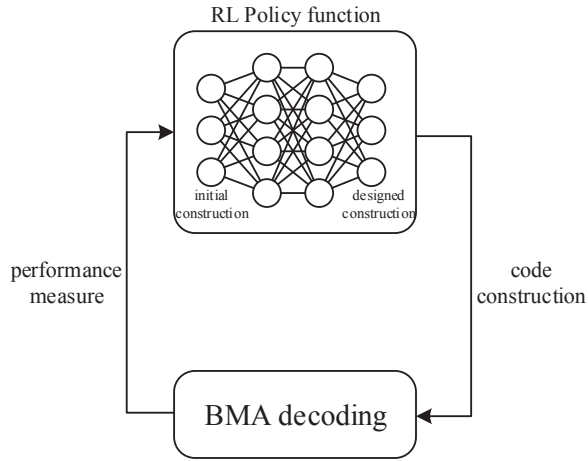}
    \caption{Framework of learning linear block codes by policy gradient.}
    \label{fig:PG:framework}
\end{figure}

The linear block code construction is modeled by a single-step MDP with a PG algorithm, as shown in Fig.~\ref{fig:PG:framework}.
The state, action and reward are introduced in section~\ref{section:learning:constructor:RL}, and detailed as follows.
\begin{itemize}
  \item The \textbf{input state} $s_0$ is a code construction defined a binary generator matrix of size $K \times N$. To impose a standard form of generator matrix, the input state is always set as $s_0=[{\bf I}_{K}, {\bf 0}]$, where ${\bf I}_{K}$ is an identity matrix of size $K \times K$ and the parity part ${\bf P}={\bf 0}$ is an all-zero matrix of size $K \times (N-K)$.
  \item A Gaussian \textbf{policy function} $\pi_{\theta_{PG}}$ is implemented by a multilayer perceptron (MLP) neural network, defined by $NeuralNet(\theta_{PG})$, with two hidden layers and sigmoid output nonlinearity:
    \begin{itemize}
      \item The hidden layer width is $2K(N-K)$, a function of code dimension and code length. Here, we set it as twice the size of the neural network output.
      \item The coefficients in the neural network are defined by $\theta_{PG}$.
      \item The output of the neural network is a real-valued matrix ${\mu}$ of size $K \times (N-K)$, which defines the policy function $\pi_{\theta_{PG}}$. It is used to determine the parity part ${\bf P}$ of the generator matrix, to be described shortly.
  \end{itemize}
  \item An \textbf{action} $a$ is sampled from the Gaussian policy function $\pi_{\theta_{PG}}$ as follows. Specifically, $a$ is a real-valued matrix of size $K \times (N-K)$, where each $a_{i,j}$ is drawn from Gaussian distribution with mean ${\mu}_{i,j}$ and variance $\sigma^2=0.1$. Then, the action $a$ is quantized to a binary-valued matrix ${\bf P}$. The probability of taking this action $a$ is recorded as $\pi'_{\theta_{PG}} (s_0,a)$.
  \item The \textbf{output state} $s_1$ is updated by $[{\bf I}_{K}, {\bf P}]$, which is the generator matrix of the constructed codes.
  \item The \textbf{reward} $R$ is defined as $-EsN0$, the required EsN0 to achieve BLER=$10^{-2}$ under BMA decoding. It is also the feedback to the policy function.
    \begin{itemize}
      \item The BLER performance of code $s_1$ is defined by $BMA(s_1, EsN0, BMA_o, BMA_s)$, where $EsN0$ is the SNR point, $BMA_o$ is the order, $BMA_s$ is the control band size. See details in~\cite{MRB:BMA}.
  \end{itemize}
\end{itemize}
The PG algorithm for linear block code construction is also described in Algorithm~\ref{alg:PG}, with parameters listed in Table~\ref{tab:PG:parameters}.

\begin{algorithm}
\begin{algorithmic}
\STATE {\bf // Initialization}:
\STATE Randomly initialize the policy function $\pi_{\theta_{PG}}$;
\STATE Set initial state $s_0=[{\bf I}_{K}, {\bf 0}]$;
\STATE {\bf // Loop}:
\WHILE 1
\STATE $(s_1, \pi'_{\theta_{PG}} (s_0,a)) \leftarrow $ Constructor($s_0$)
\STATE $R \leftarrow $ Evaluator($s_1$)
\STATE $\Delta\theta_{PG} \leftarrow SGD(\pi_{\theta_{PG}},\pi'_{\theta_{PG}} (s_0,a),R)$
\STATE ${\theta_{PG}} \leftarrow (\theta_{PG}+\Delta\theta_{PG})$
\ENDWHILE
\STATE {\bf // Constructor}:
\STATE {\bf function} $(s_1, \pi'_{\theta_{PG}} (s_0,a))$ = Constructor($s_0$)
\STATE ${\mu} \leftarrow NeuralNet(\theta_{PG})$
\FOR {$i=\{1,\cdots,K\}$}
\FOR {$j=\{1,\cdots,N-K\}$}
\STATE ${a}_{i,j} \sim {\cal N}({\mu}_{i,j}, \sigma^2)$, where $\sigma^2=0.1$;
\STATE ${\bf P}_{i,j} = ({a}_{i,j}>0.5) \ ? \ 1 : 0;$
\ENDFOR
\ENDFOR
\STATE $s_1 \leftarrow [{\bf I}_{K}, {\bf P}]$;
\STATE $\pi'_{\theta_{PG}} (s_0,a) \leftarrow f_{\cal N}(a|{\mu},\sigma^2)$;
\RETURN $s_1, \pi'_{\theta_{PG}} (s_0,a)$
\STATE {\bf end function}
\STATE {\bf // Evaluator}:
\STATE {\bf function} $R$ = Evaluator($s_1$)
\STATE Obtain $EsN0$ such that $BLER=0.01 \leftarrow BMA(s_1, EsN0, BMA_o, BMA_s)$;
\STATE $R = -EsN0$;
\RETURN $R$
\STATE {\bf end function}
\end{algorithmic}
\caption{Policy gradient based linear block codes design}
\label{alg:PG}
\end{algorithm}

\begin{table}
  \centering
  \caption{Policy Gradient algorithm parameters}
\begin{tabular}{|c|c|}
  \hline
  Parameters                                & values  \\ \hline
  Policy function hidden layer number       & $2$ \\ \hline
  Policy function hidden layers width       & $2K(N-K)$ \\ \hline
  Batch size                                & 1024 \\ \hline
  Learning rate                             & $10^{-5}$ \\ \hline
  Reward                                    & $-EsN0$ \\ \hline
  Decoder                                   & BMA \\ \hline
\end{tabular}
  \label{tab:PG:parameters}
\end{table}

\begin{figure}
\centering
    \includegraphics[width = 0.45\textwidth]{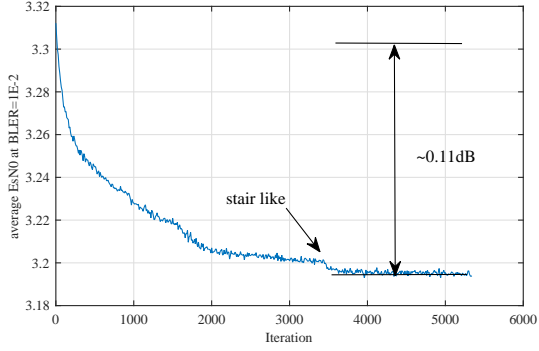}
    \caption{Evolution of the average reward per batch}
    \label{fig:PG:evolution}
\end{figure}

To optimize the policy function, the coefficients $\theta_{PG}$ in the neural network are trained by mini-batch based stochastic gradient descent (SGD) according to \eqref{equ:PG}. Here, we directly implement a method called ``AdamOptimizer'', which is integrated in Tensorflow, to update $\theta_{PG}$.
For each policy function update step, after exploring a batch of action-reward pairs, the rewards are first normalized before equation~\eqref{equ:PG} is applied.
Such processing reduces gradient estimation noise and enables a faster convergence speed.
Fig.~\ref{fig:PG:evolution} shows the evolution of average EsN0 w.r.t. the BLER of $10^{-2}$ per iteration during the learning procedure.
We observe that the average EsN0 improves in a stair-like manner.
This is because the stochastic gradient descent (SGD)-based PG algorithm tends to pull the mean of the normal distribution, from which the action is sampled, towards a local optimal point.
In the vicinity of this local optimal point, explorations are conducted randomly, which accounts for the performance fluctuations as if it has converged. Due to the high dimension of the action space, a small-probability exploration would be helpful to avoid local optimum.

\begin{figure}
\centering
    \includegraphics[width = 0.45\textwidth]{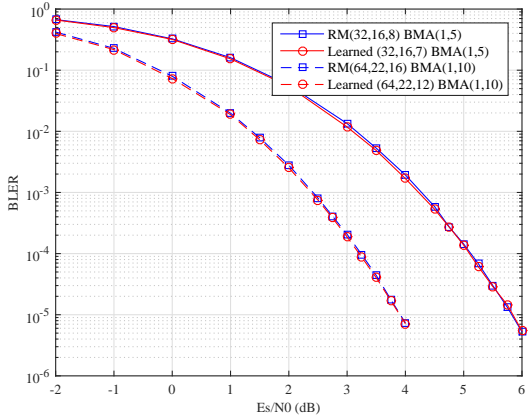}
    \caption{Performance comparison between learned linear block codes and RM codes}
    \label{fig:PG:vsRM}
\end{figure}

\begin{figure}
\centering
    \includegraphics[width = 0.45\textwidth]{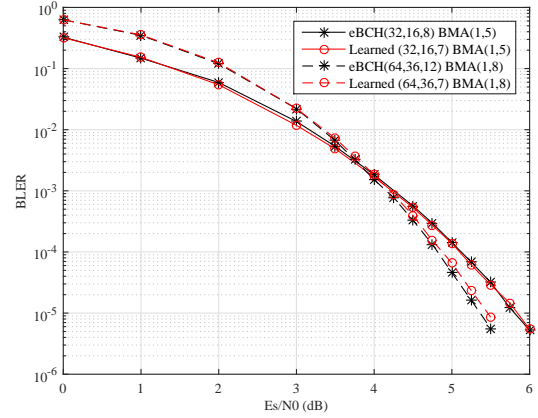}
    \caption{Performance comparison between learned linear block codes and eBCH codes}
    \label{fig:PG:vsBCH}
\end{figure}

The error correction performance comparison between the learned code constructions, RM codes and extended BCH (eBCH) codes are plotted.
In Fig.~\ref{fig:PG:vsRM}, the learned linear block codes show similar performance to RM codes for cases of ($N=32,K=16$) and ($N=64,K=22$).
In Fig.~\ref{fig:PG:vsBCH}, the learned linear block codes show similar performance to eBCH codes for cases of ($N=32,K=16$) and ($N=64,K=36$).

It is interesting that for the case of $N=32,K=16$, though the minimum code distance of the learned code ($D=7$) is smaller than that of RM code or eBCH code ($D=8$),
there is no obvious performance difference within the practical SNR range (BLER within $10^{-4} \sim 10^{-1}$).
In this SNR range for the considered case, the error correction performance of linear block codes is determined by the code distance spectrum, not only the minimum distance.

Alternatively, the design of linear block codes also can be modeled as a multi-step MDP.
For example, from an initial state, an action can be defined as determining one column (or row) of matrix ${\bf P}$ per step, or sequentially flipping each entry of matrix ${\bf P}$ per step.
Furthermore, Monte Carlo tree search (MCTS) can be incorporated into reinforcement learning to potentially enhance code performance~\cite{LDPC:WCSP}.

\subsection{Binary vector: polar codes with a fixed length and rate}
Polar codes can be defined by ${\bf c}={\bf u}{\bf G}$~\cite{Polar:Systematic}. A code construction is defined by a binary vector $\bf s$ of length $N$, in which 1 denotes an information subchannel and 0 denotes a frozen subchannel. Denote by ${\cal I}$, the support of $s$, the set of information subchannel indices. The $K$ information bits are assigned to subchannels with indices in $\cal I$, i.e., ${\bf u}_{\cal I}$. The remaining $N-K$ subchannels, constituting the frozen set $\mathcal{F}$, are selected for frozen bits (zero-valued by default). The generator matrix consists of the $K$ rows, indicated by $\cal I$, of the polar transformation matrix ${\bf G} = {\bf F}^{\otimes n}$, where ${\bf F} = \begin{bmatrix}1 & 0 \\ 1 & 1\end{bmatrix}$  is the kernel and $^\otimes$ denotes Kronecker power, and ${\bf c}$ is the codeword.

For the decoders, both SC and SCL type decoders are considered.
An SC decoder recursively computes the transition probabilities of polarized subchannels, and sequentially develops a ``path", i.e., hard decisions up to the current ($i$-th) information bits ${\bf{\hat u}} \triangleq {\hat u}_1, {\hat u}_2, \cdots, {\hat u}_i$.
At finite code length, an SCL decoder brings significant performance gain, which is briefly described below.
\begin{enumerate}
  \item run $L$ instances of SC in parallel, keep $L$ paths;
  \item extend the $L$ paths (with both $0,1$ decisions) to obtain $2L$ paths, and evaluate their path metrics (PMs);
  \item preserve the $L$ most likely paths with smallest PMs.
\end{enumerate}
Upon reaching the last bit, only one path is selected as decoding output. We consider two types of SCL decoders, characterized as follows.
\begin{itemize}
  \item SCL-PM: select the first path, i.e., with smallest PM;
  \item SCL-Genie: select the correct path, as long as it is among the $L$ surviving paths.
\end{itemize}
In practice, SCL-Genie can be implemented by CA-SCL~\cite{Polar:List_Tal,Polar:CA_List_Niu}, which selects the path that passes the CRC check. With a moderate number of CRC bits (e.g., 24), CA-SCL yields almost identical performance to SCL-Genie.

\begin{figure}
\centering
    \includegraphics[width = 0.5\textwidth]{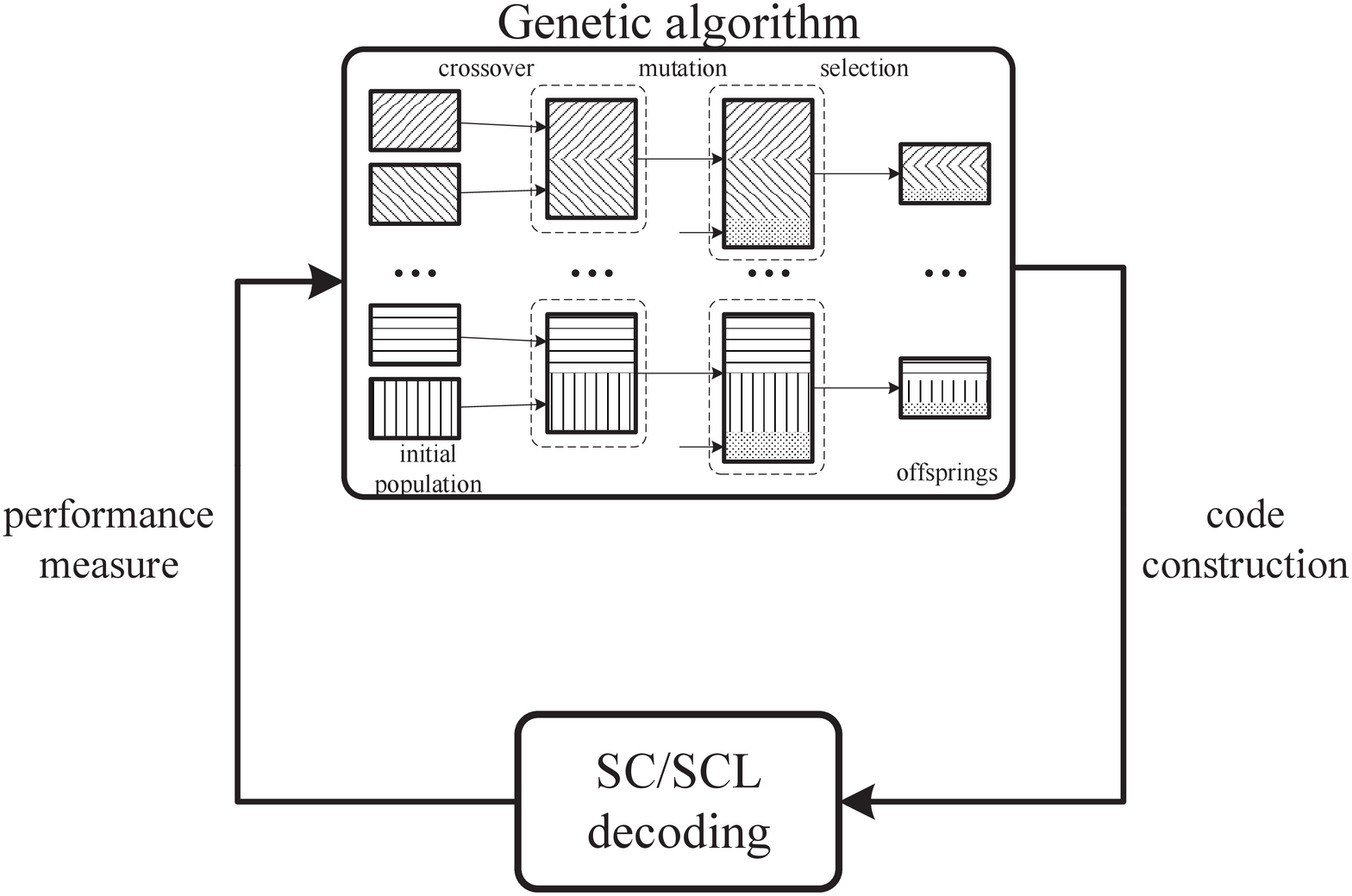}
    \caption{Framework of learning polar codes by genetic algorithm.}
    \label{fig:GA:framework}
\end{figure}

Genetic algorithm is applied to construct polar codes for various types of decoders. We observe that the information subchannels in a code construction play the same role of chromosomes in genetic algorithm, because they both individually and collaboratively contribute to the fitness of a candidate solution. The key insight is that good code constructions are constituted by good subchannels. Therefore, a pair of good parent code constructions is likely to produce good offspring code constructions. This suggests that a genetic algorithm may ultimately converge to a good code construction. The framework is shown in Fig.~\ref{fig:GA:framework}, and the algorithm is detailed below.

During the initialization of the genetic algorithm, the population (code constructions) are randomly generated. Specifically, the information subchannel set $\cal I$ is randomly and uniformly selected from $\{1,\cdots,N\}$ for each code construction \emph{without} given prior knowledge about existing polar code construction techniques such as Bhattacharyya~\cite{Polar:Arikan} and DE/GA~\cite{Polar:DE3_Tal,Polar:GA_Trifonov}. The purpose is to test whether the genetic algorithm can learn a good code construction without this expert knowledge.

A population of $M$ code constructions are randomly initialized and sorted according to ascending BLER performance $BLER_{{\cal I}_i}\triangleq{BLER_1\times BLER_2}$, which is defined as the product of BLERs at two SNR points. Typically, $M$ should be sufficiently large to store all the good chromosomes (subchannel indices) to ensure an efficient convergence. A polar decoder, denoted by $BLER_x \leftarrow PolarDecoder({\cal I}_i,SNR_x)$, returns the BLER performance of the constructed codes at a specified SNR point. At least 1000 block error events are collected to ensure an accurate estimate.

\begin{figure*}
\centering
    \includegraphics[width = 0.9\textwidth]{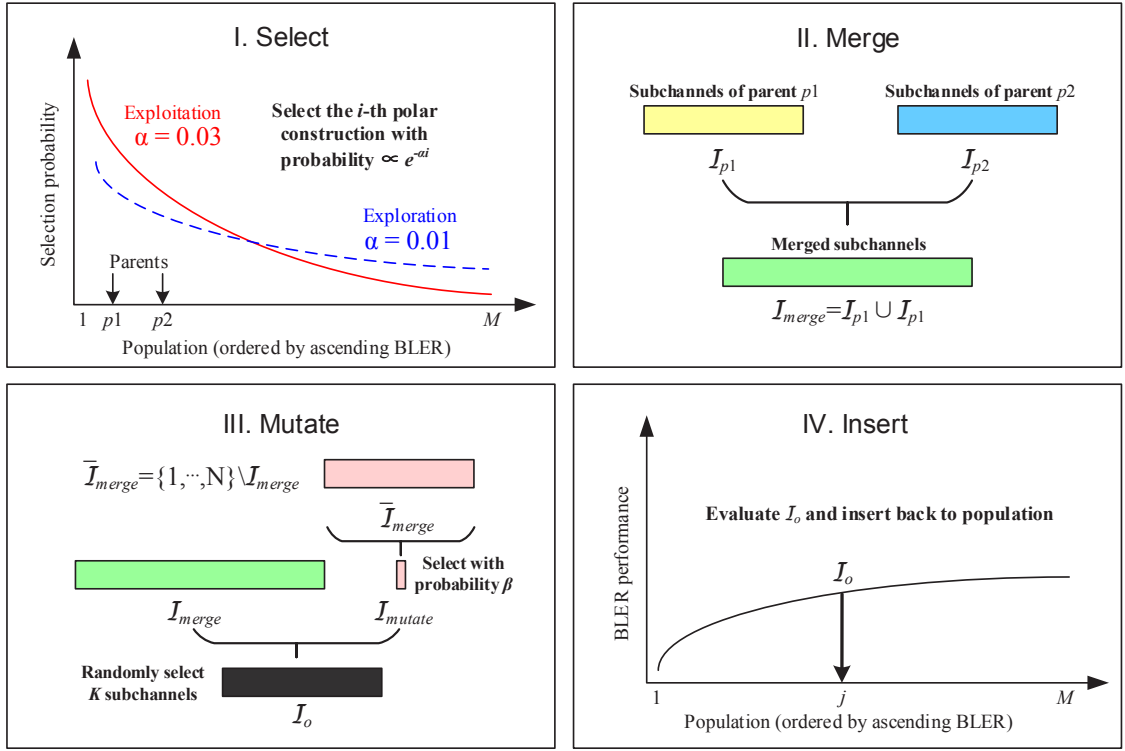}
    \caption{Genetic algorithm for polar code construction in four steps.}
    \label{fig:GA:steps}
\end{figure*}
After initialization, the algorithm enters a loop consisting of four steps, as shown in Fig.~\ref{fig:GA:steps}.
\begin{enumerate}
  \item \textbf{Select} parents ${\cal I}_{p1}, {\cal I}_{p2}$ from population. The $i$-th code construction is selected according to a probability distribution $e^{-\alpha i} / \sum_{j=1}^M e^{-\alpha j}$ (normalized), where $\alpha$ is called the \emph{sample focus}. Specifically, the corresponding cumulative distribution function \texttt{cdf} is sampled uniformly at random to determine the index $p$ to be selected. This is implemented in Matlab with a command: \texttt{[$\sim$,p]=histc(rand,cdf)}. In this way, a better code construction will be selected with a higher probability. Two distinct code constructions, denoted by their information subchannel sets ${\cal I}_{p1}$ and ${\cal I}_{p2}$, are selected as parents. By adjusting the parameter $\alpha$, we can tradeoff between exploitation (a larger $\alpha$) and exploration (a smaller $\alpha$).
  \item \textbf{Merge} subchannels from parents by ${\cal I}_{merge} = {\cal I}_{p1} \cup {\cal I}_{p2}$. Note that $|{\cal I}_{p1}| = |{\cal I}_{p2}| = K$ and their union set contains more than $K$ subchannels. This ensures that an offspring code construction contains the subchannels from both parents. In fact, it implements the ``crossover'' function described in Section~\ref{section:learning:constructor:genetic}.
  \item \textbf{Mutate} the code construction by including a few mutated subchannels ${\cal I}_{mutate}$ from ${\overline{\cal I}}_{merge} = \{1,\cdots,N\}\backslash {\cal I}_{merge}$, that is, the remaining subchannels. Specifically, $\lfloor \beta \times |{\cal I}_{merge}|\rceil$ indices are randomly and uniformly selected from ${\overline{\cal I}}_{merge}$, where $\beta$ is called the \emph{mutation rate}. Finally, the $K$ information subchannels of the offspring ${\cal I}_o$ are randomly and uniformly selected from ${\cal I}_{merge} \cup {\cal I}_{mutate}$. Note that the offspring may or may not contain the mutated subchannels depending on whether those in ${\cal I}_{mutate}$ are finally selected into ${\cal I}_o$. The mutation rate $\beta$ provides a way to control exploration. The larger $\beta$ is, the more likely a mutated subchannel is included in the offspring.
  \item \textbf{Insert} the offspring ${\cal I}_o$ back to population according to ascending BLER performance. If its BLER performance is worse than all existing ones in the population, it is simply discarded.
\end{enumerate}

The while loop can be terminated after a maximum number of iterations $T_{\max}$ is reached, or a desired performance is obtained by the best code construction in the population. The algorithm is described in Algorithm~\ref{alg:GeneticPolar}.

\begin{algorithm}
\begin{algorithmic}
\STATE {\bf // Parameters}:
\STATE population size $M=1000$, sample focus $\alpha=0.03$, mutation rate $\beta=0.01$, $SNR_1$ and $SNR_2$;
\STATE {\bf // Initialization}:
\FOR {$i = \{1,2,\cdots,M\}$}
\STATE ${\cal I}_i \leftarrow$ randomly and uniformly select $K$ indices from $\{1,\cdots,N\}$
\STATE $BLER_{{\cal I}_i} \leftarrow {BLER_1\times BLER_2}$, where $BLER_x \leftarrow PolarDecoder({\cal I}_i,SNR_x)$
\ENDFOR
\STATE Sort $\{{\cal I}_1,\cdots,{\cal I}_M\}$ such that $BLER_{{\cal I}_i}<BLER_{{\cal I}_j}, \quad \forall \ i<j$;
\STATE $t=0$
\WHILE {$t<T_{\max}$}
\STATE $t \leftarrow t+1$
\STATE {\bf // Select}:
\FOR {$k = \{1,2\}$}
\STATE \texttt{cdf} $\leftarrow$ \texttt{pdf}$=e^{-\alpha i} / \sum_{j=1}^M e^{-\alpha j}$
\STATE Uniformly sample the \texttt{cdf} to determine $p_k$: \texttt{[$\sim$,$p_k$]=histc(rand,cdf)}
\STATE ${\cal I}_{p_k} \leftarrow $ select ${\cal I}_{p_k}$ from $\{{\cal I}_1,\cdots,{\cal I}_M\}$
\ENDFOR
\STATE {\bf // Merge}:
\STATE ${\cal I}_{merge} = {\cal I}_{p1} \cup {\cal I}_{p2}$
\STATE {\bf // Mutate}:
\STATE ${\overline{\cal I}}_{merge} = \{1,\cdots,N\}\backslash {\cal I}_{merge}$
\STATE ${\cal I}_{mutate} \leftarrow$ randomly and uniformly select $\lfloor \beta \times |{\cal I}_{merge}|\rceil$ indices from ${\overline{\cal I}}_{merge}$
\STATE ${\cal I}_{o} \leftarrow$ randomly and uniformly select $K$ indices from ${\cal I}_{merge} \cup {\cal I}_{mutate}$
\STATE {\bf // Insert}:
\STATE $\{{\cal I}_1,\cdots,{\cal I}_M,{\cal I}_{M+1}\} = \{{\cal I}_1,\cdots,{\cal I}_M\}\cup{\cal I}_{o}$ where ${\cal I}_{M+1} = {\cal I}_{o}$
\STATE $BLER_{{\cal I}_o} \leftarrow {BLER_1\times BLER_2}$, where $BLER_x \leftarrow PolarDecoder({\cal I}_o,SNR_x)$
\STATE Sort $\{{\cal I}_1,\cdots,{\cal I}_{M+1}\}$ such that $BLER_{{\cal I}_i}<BLER_{{\cal I}_j}, \quad \forall \ 1\leq i<j \leq M+1$
\ENDWHILE
\STATE {\bf // Output}:
\STATE ${\cal I}_1$ and $BLER_{{\cal I}_1}$
\end{algorithmic}
\caption{Genetic algorithm based polar code design}
\label{alg:GeneticPolar}
\end{algorithm}

We compare the learned code constructions with the baseline schemes below:
\begin{itemize}
  \item ${\cal I}_{DE/GA}$, a close-to-optimal construction under SC decoding that is obtained via GA~\cite{Polar:GA_Trifonov}: we search for a design SNR (starting from -20dB, increasing with step size 0.25dB) to obtain a construction that requires the lowest SNR to achieve BLER=$10^{-2}$.
  \item ${\cal I}_{PW}$, an SNR-independent construction obtained by PW~\cite{Polar:PW,Polar:betaExpansion}.
  \item ${\cal I}_{RM-Polar}$, a heuristic construction~\cite{Polar:RM_Polar_Li} that yields better performance under SCL-PM decoder. The only difference from~\cite{Polar:RM_Polar_Li} is that we use PW~\cite{Polar:PW,Polar:betaExpansion} as the reliability metric to make the design SNR-independent.
\end{itemize}

We use them as baselines, and observe the learning process through two metrics: (1) the BLER performance of the learned codes at two SNR points, which corresponds to the required SNRs to achieve BLER=$10^{-1}$ and $10^{-2}$ for the baseline scheme (2) the difference between the learned information subchannels ${\cal I}_{learned}$ and those of baseline schemes.

To demonstrate the effectiveness of genetic algorithm, we record the first time (iteration number) a code construction converges to the existing optimal construction, i.e., ${\cal I}_{learned}={\cal I}_{DE/GA}$ under SC decoding. The results for different code lengths are shown in Table~\ref{tab:genetic:SC_convergence}. Note that a brute-force search would check all $nchoosek(N,K)$ possible candidates, which is prohibitively costly as shown in Table~\ref{tab:genetic:SC_convergence}. By contrast, a reasonable converging time is observed with the genetic algorithm at medium code length. There is no big difference in terms of learning efficiency between the SC and SCL decoders. For SC decoder, there exists already an explicit optimum way of constructing polar codes and the learned construction could not outperform that. However, the optimal code constructions for both SCL-PM and SCL-Genie are still open problems. The above results imply that we can apply genetic algorithm to SCL decoders to obtain good performance within a reasonable time.

\begin{table}
  \centering
  \caption{Convergence time for learning process}
\begin{tabular}{|c|c|c|c|c|}
  \hline
  $N$ & $K$ & design EsN0 (dB) & nchoosek($N,K$) & \# iterations \\ \hline
  16 & 8 & 4.50 & 12870 & 126 \\
  32 & 16 & 4.00 & $6.0 \times 10^9$ & 623 \\
  64 & 32 & 3.75 & $1.8 \times 10^{19}$ & 2100 \\
  128 & 64 & 3.50 & $2.4 \times 10^{37}$ & 5342 \\
  256 & 128 & 3.25 & $5.8 \times 10^{75}$ & 19760 \\
  512 & 256 & 3.00 & $4.7 \times 10^{152}$ & 56190 \\
  \hline
\end{tabular}
  \label{tab:genetic:SC_convergence}
\end{table}

\begin{figure}
\centering
    \includegraphics[width = 0.45\textwidth]{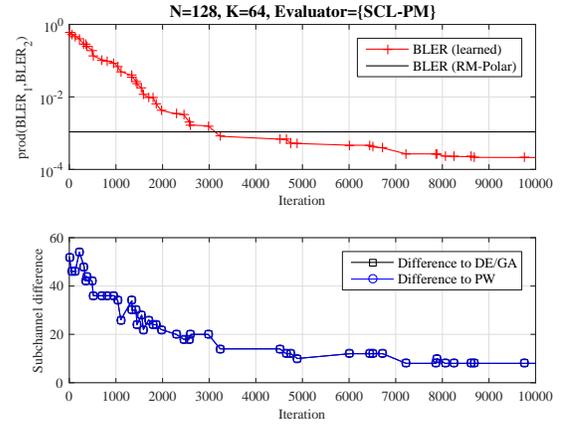}
    \caption{Evolution of learned polar code constructions (information subchannels) and BLER performance under SCL-PM ($L=8$) decoder. The performance is defined as the product of BLERs measured at EsN0=[1.74, 2.76]dB. The DE/GA is constructed with design EsN0 at 3.5dB. In this case, the PW construction coincides with DE/GA.}
    \label{fig:genetic:SCL_PM_N128K64_Evolution}
\end{figure}

\begin{figure}
\centering
    \includegraphics[width = 0.45\textwidth]{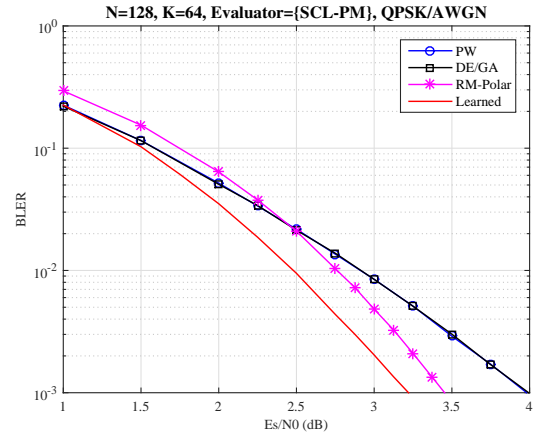}
    \caption{BLER comparison between learned polar code constructions (information subchannels) and \{DE/GA, PW, RM-Polar\} under SCL-PM ($L=8$) decoder. The DE/GA is constructed with design EsN0 at 3.5dB.}
    \label{fig:genetic:SCL_PM_N128K64_BLER}
\end{figure}

We first consider SCL-PM with $L=8$. The learning process for $N=128, K=64$ is shown in Fig.~\ref{fig:genetic:SCL_PM_N128K64_Evolution}. Note that DE/GA is derived assuming an SC decoder. We adopt RM-Polar~\cite{Polar:RM_Polar_Li} to obtain the baseline performance. In this case, the minimum distance is 16 for RM-Polar and 8 for DE/GA and PW. In this case, the PW construction coincides with DE/GA, i.e., ${\cal I}_{PW}={\cal I}_{DE/GA}$. In the upper subfigure, the performance (product of BLERs measured at EsN0=[1.74, 2.76]dB) of the learned codes quickly converges and outperforms that of the RM-polar codes at iteration 3100. In the lower subfigure, unlike the case of the SC decoder, the difference between ${\cal I}_{learned}$ and ${\cal I}_{DE/GA}$ (design EsN0 at 3.5dB) stops to decrease after reaching 8. The learned codes outperform DE/GA by 0.8 dB and RM-polar by 0.2 dB at BLER=$10^{-3}$, as shown in Fig.~\ref{fig:genetic:SCL_PM_N128K64_BLER}. These results demonstrate that, in the current absence of an optimal coding-theoretic solution, learning algorithms can potentially play important roles in code construction.


\begin{figure}
\centering
    \includegraphics[width = 0.45\textwidth]{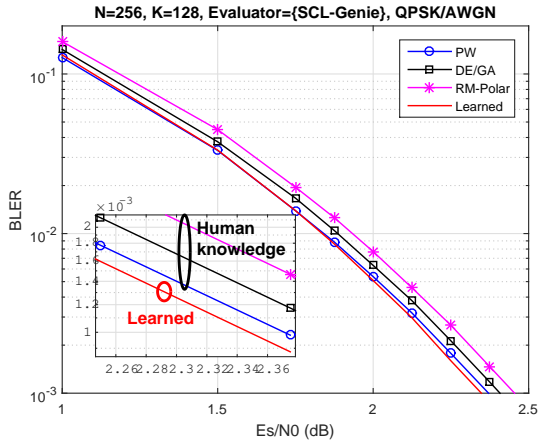}
    \caption{BLER comparison between learned polar code constructions (information subchannels) and \{DE/GA, PW\} under SCL-Genie ($L=8$) decoder. The DE/GA is constructed with design EsN0 at 3.25dB.}
    \label{fig:genetic:SCL_Genie_N256K128_BLER}
\end{figure}

The same observation holds for SCL-Genie, where the optimum construction is also unknown. The BLER curves for $N=256, K=128$ under SCL-Genie with $L=8$ is shown in Fig.~\ref{fig:genetic:SCL_Genie_N256K128_BLER}. We aim at DE/GA (design EsN0 at 3.25dB) and PW constructions as the baseline schemes since they perform better than RM-Polar under SCL-Genie. In this case, the minimum distance is 16 for RM-Polar and 8 for DE/GA and PW. In the genetic algorithm, the evaluator measures performance as product of BLERs at EsN0=[1.17, 1.88]dB. As seen, the learned codes perform better than those generated by DE/GA and PW, with a slightly better slope. However, the performance gain (0.06 dB between learned codes and DE/GA) is much smaller than the case with SCL-PM.

\begin{figure}
\centering
    \includegraphics[width = 0.45\textwidth]{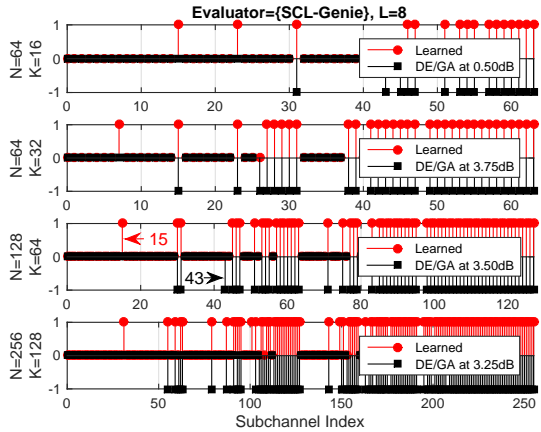}
    \caption{Difference of information subchannels between learned constructions (${\cal I}_{learned}$) and DE/GA (${\cal I}_{DE/GA}$, design EsN0 labeled in each subfigure) under SCL-Genie ($L=8$) decoder.}
    \label{fig:genetic:SCL_Genie_L8_InfoSetDiff}
\end{figure}

It is of interest to observe the difference between the learned information subchannels and those selected by DE/GA. In Fig.~\ref{fig:genetic:SCL_Genie_L8_InfoSetDiff}, the subchannel differences between the learned construction and DE/GA under various code lengths and rates are plotted, where the positive positions are ${\cal I}_{learned}$ and negative ones are ${\cal I}_{DE/GA}$ (design EsN0 are labeled in each subfigure). The evaluator is SCL-Genie with $L=8$. The first observation is that all the learned constructions prefer subchannels with smaller indices. A close look would reveal that the learned construction may violate the universal partial order (UPO)~\cite{Polar:UPO_Sch¨¹rch,Polar:UPO_Bardet}. For the case of $N=128, K=64$ (the genetic algorithm evaluates performance as product of BLERs at EsN0=[0.94, 1.89]dB.), the only difference is that ${\cal I}_{learned}$ preferred the 15-th subchannel over the 43-th. It is easy to verify that this choice violates the UPO because $15=[0,0,0,1,1,1,1]$ and $43=[0,1,0,1,0,1,1]$ in binary form. Note that UPO applies in theory to the SC decoder, rather than the SCL decoder.

In this subsection, we demonstrate that good polar codes can be learned for SC, SCL-PM and SCL-Genie decoders using genetic algorithm. For SCL decoders, the learned codes may even outperform existing ones. Note that a recent independent work \cite{Polar:GeneAlg} proposed very similar approaches to this subsection. The main difference is that prior knowledge such as Bhattacharyya construction \cite{Polar:Arikan} and RM-Polar construction \cite{Polar:RM_Polar_Li} is utilized by \cite{Polar:GeneAlg} during the population initialization to speed up the learning process, while no such knowledge was exploited in our work. The detailed differences are summarized below:
\begin{itemize}
  \item The initialization of learning algorithms is different. The work \cite{Polar:GeneAlg} initializes the (code constructions) population based on the Bhattacharyya construction obtained for BECs with various erasure probabilities, and RM-Polar construction. In this way, the convergence time is significantly reduced. In our experiment, we randomly initialize the population to test whether the genetic algorithm can learn a good code construction without these prior knowledge. One of our motivation is to answer Q1 in section~\ref{section:intro:AI}.
  \item The new population is generated in a different way. In \cite{Polar:GeneAlg}, the $T$ best members in the population is always secured in the next-iteration population, the next $nchoosek(T,2)$ members are generated through crossover of the $T$ best members, and the final $T$ members are generated by mutation. In this work, we select a pair of parents according to certain sample probability (a better member is selected with higher probability), which provides a flexible tradeoff between exploration and exploitation.
  \item The crossover operation is not exactly the same. In \cite{Polar:GeneAlg}, the offspring takes half of the subchannels from each of the parents. In our work, the subchannels of both parents are first merged, followed a mutation step that includes a few mutated subchannels. The resulting offspring randomly takes $K$ subchannels from among these subchannels.
  \item The cost function of \cite{Polar:GeneAlg} is the error-rate at a single SNR point, whereas our work allows to choose a set of SNR points for potential benefit of controlling the slope of error rate.
  \item The work \cite{Polar:GeneAlg} also tried belief propagation decoder, whereas we focus on the SC-based decoders due to their superior performances.
\end{itemize}

\subsection{Nested representation: polar codes within a range of code rates}
We evaluate another type of polar code constructions with nested property, which bears practical significance due to its description and implementation simplicity. Polar codes for code length $N$ and dimension range $[K_l, K_h]$ are defined as a reliability ordered sequence of length $N$. The corresponding polar code can be determined by reading out the first (or last) $K$ entries from the ordered sequence to form the information position set.

The code design procedure is modeled by a multi-step MDP.
Specifically for each design step, with a given $(N,K)$ polar code (current state), a new subchannel (action) is selected, to get the $(N,K+1)$ polar code (an updated state). The reliability ordered sequence is constructed by sequentially appending the actions to the end of initial polar code construction.

Note that the optimal constructions for $(N,K)$ polar codes and $(N,K+1)$ polar codes may not be nested, i.e., the information set of $(N,K)$ polar codes may not be a subset of the information set of $(N,K+1)$ polar codes. As a result, the optimal constructions for different $(N,K)$ polar codes with do not necessarily constitute a nested sequence. In other words, the performance of some $(N,K)$ codes needs to be compromised for the nested property. Therefore, during the construction of the reliability ordered sequence, a tradeoff exists between the short-term reward (from the next state construction) and long-term reward (from the construction of a state that is a few steps away). The problem of maximizing the total reward, which consists of both short-term reward and long-term reward, can be naturally solved by the A2C algorithm.

\begin{figure}
\centering
    \includegraphics[width = 0.5\textwidth]{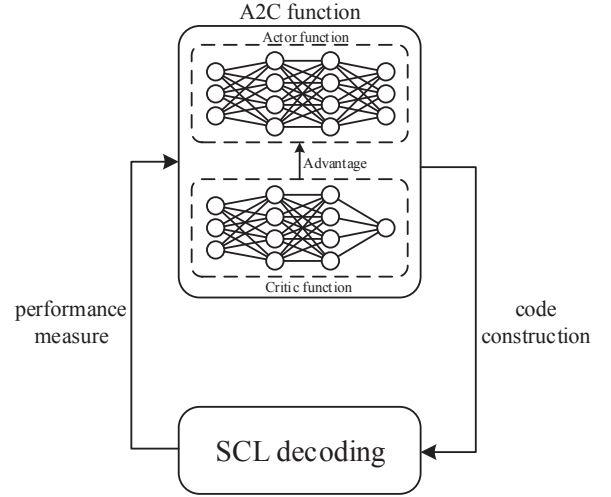}
    \caption{Framework of learning polar codes by advantage actor critic algorithm.}
    \label{fig:A2C:framework}
\end{figure}

The framework is shown in Fig.~\ref{fig:A2C:framework}.
The state, action and reward are introduced in section~\ref{section:learning:constructor:RL}, and detailed as follows.
\begin{itemize}
  \item The \textbf{input state} $s_t$ is a code construction defined a binary vector of length $N$, in which 1 denotes an information subchannel and 0 denotes a frozen subchannel. Denote by ${\cal I}_{s_t}$, the support of $s_t$, the set of information subchannel indices.
  \item Both the \textbf{actor function} $\pi_{\theta_{A}}(s_t)$ and \textbf{critic function} $V_{\theta_{C}}(s_t)$ are implemented by MLP neural networks, defined by $NeuralNet(\theta_{A})$ and $NeuralNet(\theta_{C})$, with two hidden layers and sigmoid output nonlinearity:
    \begin{itemize}
      \item The input to both of the actor and critic functions is the state $s_t$ (code construction).
      \item The hidden layer width is $4N$.
      \item The coefficients in the neural network are defined by $\theta_{A}$ and $\theta_{C}$, respectively.
      \item The output of $NeuralNet(\theta_{A})$ is a probability mass function $\pi_{\theta_{A}}(s_t)$ of all possible action $a_t \in \{1,\cdots,N\}$ taken at state $s_t$.
      \item The output of $NeuralNet(\theta_{C})$ is a state value estimation $V_{\theta_{C}}(s_t)$ for the state $s_t$.
  \end{itemize}
  \item An \textbf{action} $a_t$ denotes the next information subchannel position to be selected, which is sampled according to the probability mass function $\pi_{\theta_{A}}(s_t)$.
  \item The \textbf{output state} $s_{t+1}$ is updated by setting the $a_t$-th position in $s_t$ to 1, i.e., $s_{t+1} \leftarrow s_{t}[a_t]=1$.
  \item The \textbf{reward} $R_{t}$ is defined as $\log(BLER)$ at a given SNR point. It is also the feedback to the A2C function block.
    \begin{itemize}
      \item The BLER performance of code construction $s_{t+1}$ is evaluated by the SCL decoding (either SCL-PM or SCL-Genie decoding).
  \end{itemize}
\end{itemize}
The A2C based polar code reliability ordered sequence design is also described in Algorithm~\ref{alg:A2C}, with parameters listed in Table~\ref{tab:A2C:parameters}.

To optimize the A2C function block, the coefficients $\theta_{C}$ and $\theta_{A}$ in the neural networks are trained by mini-batch based SGD according to \eqref{equ:A2C_C} and \eqref{equ:A2C_A}, respectively. Specifically, the next state $s_{t+1}$ and the reward $R_{t}$ are feed back to A2C function block. Note that the advantage value $Adv(s_t,s_{t+1},R)$ in \eqref{equ:A2C_C} and \eqref{equ:A2C_A} is calculated according to \eqref{equ:A2C_Adv}, which requires critic function and the current reward. The ``AdamOptimizer'' in Tensorflow is applied to update $\theta_{C}$ and $\theta_{A}$.

\begin{algorithm}
\begin{algorithmic}
\STATE Initialize the coefficients $\theta_{A}$ and $\theta_{C}$ randomly;
  \WHILE 1
    \STATE $s_0 = zeros(1,N)$;
    \STATE $s_0 \leftarrow s_0[N-1,N-2,\cdots,N-K_l+1]=1$;
    \FOR{$t=0$ to $(K_h-K_l)$}
      \STATE $V_{\theta_{C}} \leftarrow NeuralNet(\theta_{C})$
      \STATE $\pi_{\theta_{A}} \leftarrow NeuralNet(\theta_{A})$
      \STATE $a_t \sim \pi_{\theta_{A}}(s_t)$;
      \STATE $s_{t+1} \leftarrow s_{t}[a_t]=1$;
      \STATE ${\cal I}_{s_{t+1}} \leftarrow support(s_{t+1})$;
      \STATE $BLER \leftarrow PolarDecoder({\cal I}_{s_{t+1}},EsN0)$;
      \STATE $R_{t}=\log(BLER)$;
      \STATE $\Delta\theta_{C} \leftarrow SGD(V_{\theta_{C}}, s_{t}, s_{t+1}, R_t)$;
      \STATE $\theta_{C} \leftarrow (\theta_{C}+\Delta\theta_{C})$
      \STATE $\Delta\theta_{A} \leftarrow SGD(\pi_{\theta_{A}}, s_{t}, a_t, s_{t+1}, R_t)$;
      \STATE $\theta_{A} \leftarrow (\theta_{A}+\Delta\theta_{A})$
    \ENDFOR
    \STATE Sequence is $\{N-1,N-2,\cdots,N-K_l+1,a_0, a_1, \cdots, a_{K_h-K_l}\}$.
  \ENDWHILE
\end{algorithmic}
\caption{A2C based polar code reliability ordered sequence design}
\label{alg:A2C}
\end{algorithm}

\begin{table}
  \centering
  \caption{A2C algorithm parameters}
\begin{tabular}{|c|c|}
  \hline
  Parameters                            & Values  \\ \hline
  Actor function hidden layer number    & $2$ \\ \hline
  Actor function hidden layer width     & $4N$ \\ \hline
  Critic function hidden layer number   & $2$ \\ \hline
  Critic function hidden layer width    & $4N$ \\ \hline
  Batch size                            & 32 \\ \hline
  Actor learning rate                   & $1.0 \times 10^{-3}$ \\ \hline
  Critic learning rate                  & $2.0 \times 10^{-3}$ \\ \hline
  Reward discount factor                & $0.2$ \\ \hline
  Reward                                & $\log(BLER)$ \\ \hline
  Decoder                               & SCL-PM and SCL-Genie with $L=8$ \\ \hline
\end{tabular}
  \label{tab:A2C:parameters}
\end{table}

As an example, a reliability ordered sequence of length $N=64$ and dimension range $K\in [4,63]$ is constructed by A2C.
To ensure a fair comparison, the design EsN0 for the DE/GA polar codes are always optimized to obtain a construction that requires the lowest SNR to achieve BLER=$10^{-2}$.
The constructed codes are tested under the required SNRs to achieve BLER=$10^{-2}$ for the optimized DE/GA construction with the same $K$.

\begin{figure}
\centering
    \includegraphics[width = 0.45\textwidth]{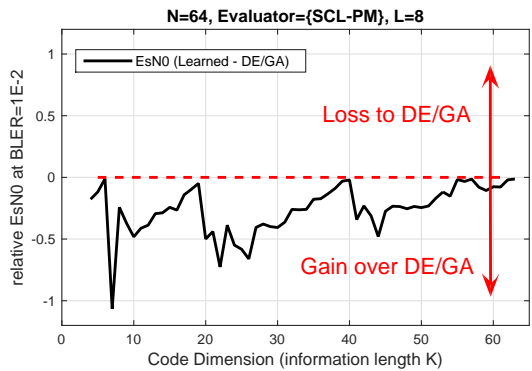}
    \caption{Relative performance between polar codes constructed by reinforcement learning and DE/GA under SCL-PM}
    \label{fig:Polar_RelativeEsN0_SCL_PM}
\end{figure}

In the first experiment, the code performance evaluator deploys SCL-PM with $L=8$. Fig.~\ref{fig:Polar_RelativeEsN0_SCL_PM} compares the relative SNR to achieve a target BLER level of $10^{-2}$.
A better overall performance can be observed for the learned polar codes, with the largest gain over 0.5 dB over DE/GA. There is almost no loss cases.

\begin{figure}
\centering
    \includegraphics[width = 0.45\textwidth]{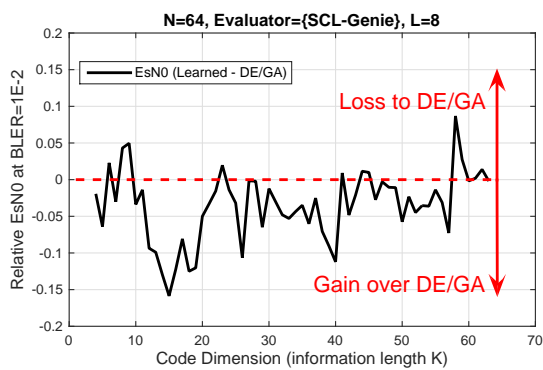}
    \caption{Relative performance between polar codes constructed by reinforcement learning and DE/GA under SCL-Genie}
    \label{fig:Polar_RelativeEsN0}
\end{figure}

In the second experiment, the code performance evaluator deploys SCL-Genie with $L=8$. Fig.~\ref{fig:Polar_RelativeEsN0} compares the SNR required to achieve a target BLER level of $10^{-2}$.
A slightly better overall performance can be observed for the learned polar codes, with the largest gain of 0.16 dB at $K=15$ over DE/GA, and the largest loss of 0.09 dB at $K=58$.

Some discussions on why good nested polar codes can be learned using A2C algorithm are as follows. Compared with the fixed $(N,K)$ case, the nested code constraint makes the optimization complicated for classical coding theory. Because there is no theory to optimize the overall performance within a range of code rates. The problem of selecting subchannels in a nested manner is essentially a multi-step decision problem. It naturally fits into the multi-step MDP model and the problem can be solved by existing RL algorithms.

\section{Conclusion and Future works}\label{section:conclusion}
In this paper, we tried to design error correction codes with AI techniques.
We employed a constructor-evaluator framework, in which the code constructor is realized by AI algorithms whose target function depends on only performance metric feedback from the code evaluator.
The implementation of the code constructor is illustrated (e.g., by reinforcement learning and genetic evolution) and the flexibility of the code evaluator is analyzed.

We have provided three detailed AI-driven code construction algorithms for three types of code representations.
In essence, the framework is able to iteratively refine code construction without being taught explicit knowledge in coding theory.
For proof of concept, we show that, for linear block codes and polar codes in our examples, the learned codes can achieve a comparable performance to the state-of-the-art ones.
For certain cases, the learned codes may even outperform existing ones.

For future works, both the constructor and evaluator design need to be explored to either solve more general problems or further improve the efficiency. Some code construction problems that were intractable under classical coding theoretic approaches may be revisited using AI approaches. Moreover, more realistic settings such as online code construction should be studied.

\ifCLASSOPTIONcaptionsoff
  \newpage
\fi

\begin{IEEEbiography}[{\includegraphics[width=1in,height=1.25in,clip,keepaspectratio]{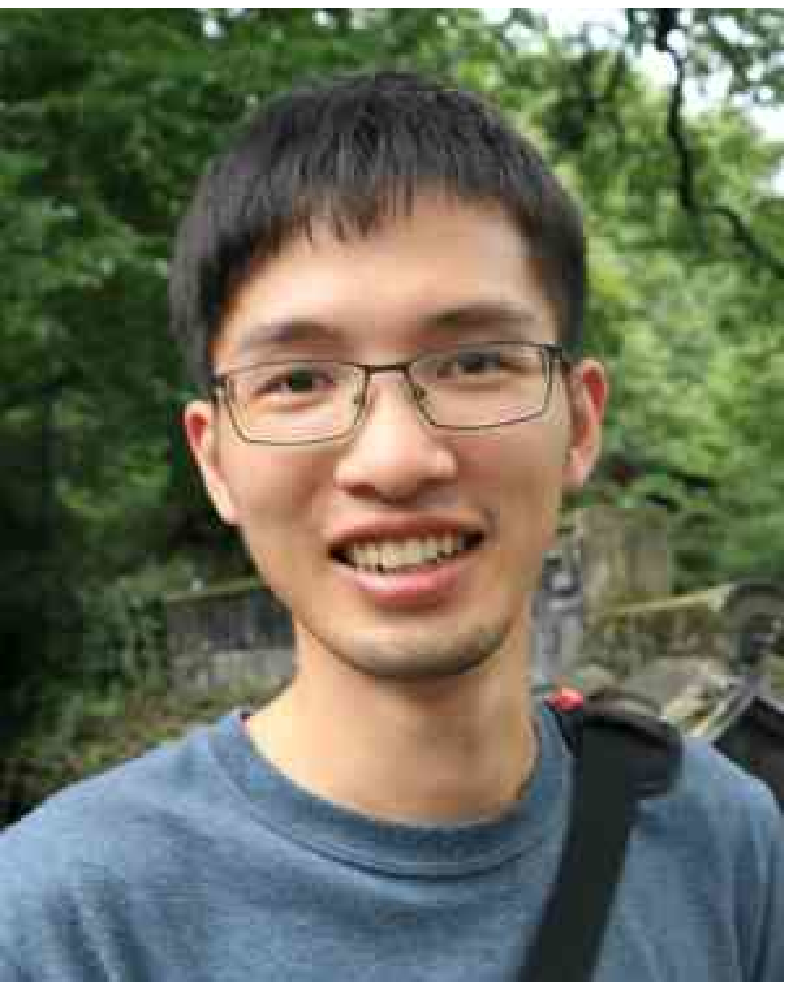}}]{Lingchen~Huang}
is a research engineer at Huawei Technologies Co., Ltd. His current research interests are channel coding schemes with focus on algorithm design and hardware implementations.
\end{IEEEbiography}

\begin{IEEEbiography}[{\includegraphics[width=1in,height=1.25in,clip,keepaspectratio]{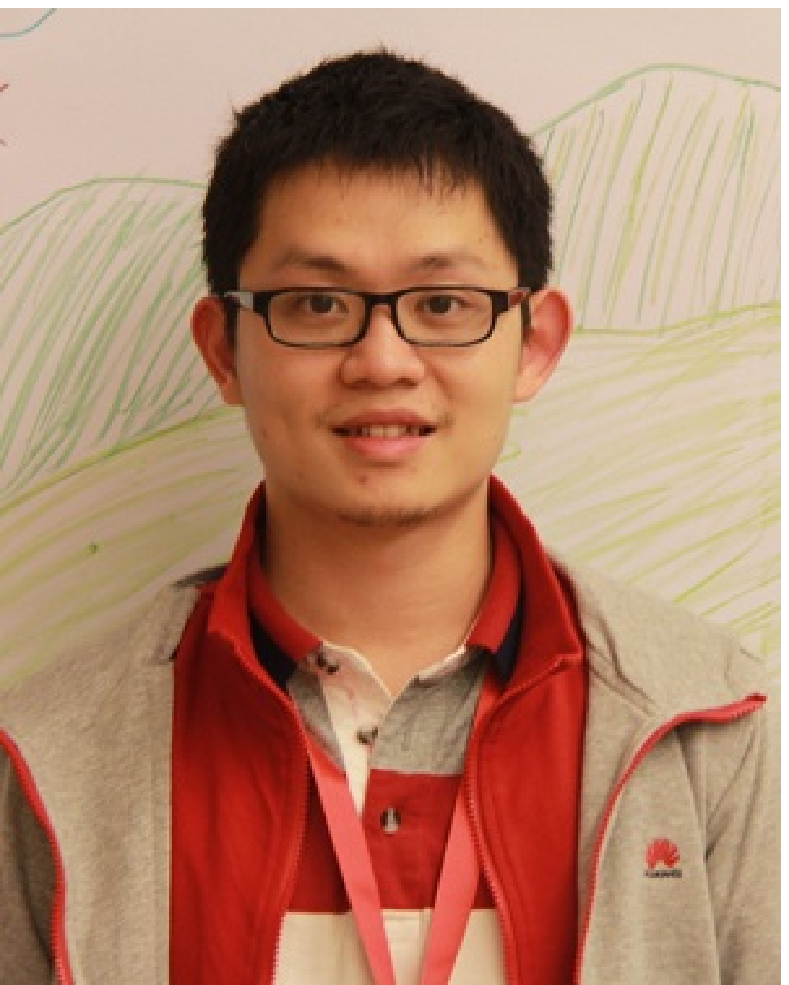}}]{Huazi~Zhang}
is a research engineer at Huawei Technologies Co., Ltd. His current research interests are channel coding schemes with focus on algorithm design and hardware implementations.
\end{IEEEbiography}

\begin{IEEEbiography}[{\includegraphics[width=1in,height=1.25in,clip,keepaspectratio]{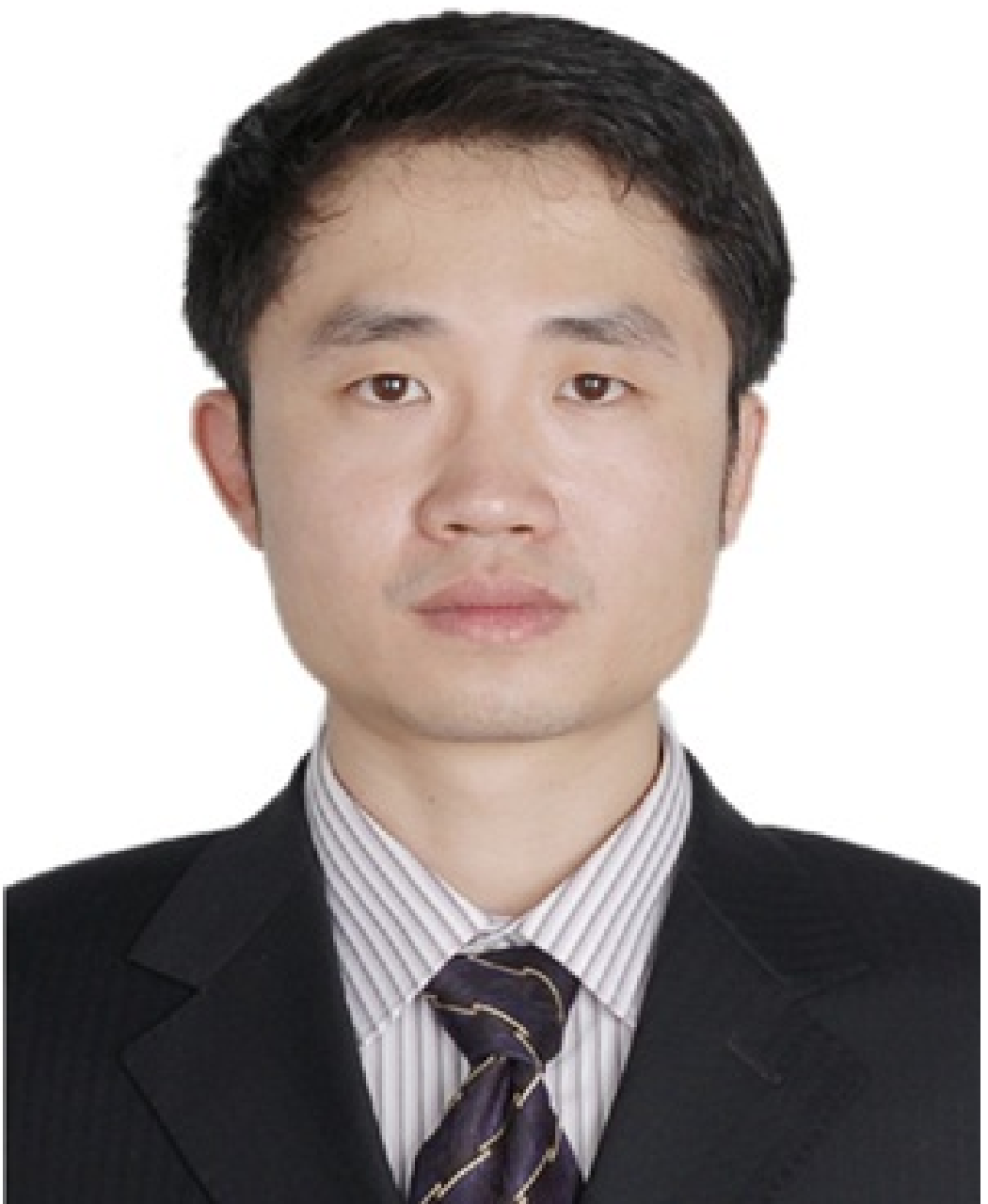}}]{Rong Li}
is a research expert at Huawei Technologies Co., Ltd. His current research interests are channel coding schemes with focus on algorithm design and hardware implementations. He has been the Technical Leader on Huawei 5G air interface design focusing mainly on channel coding.
\end{IEEEbiography}

\begin{IEEEbiography}[{\includegraphics[width=1in,height=1.25in,clip,keepaspectratio]{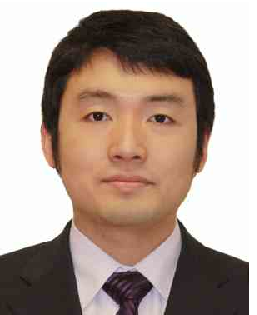}}]{Yiqun~Ge}
is a distinguished research engineer at Huawei Technologies Co., Ltd. His research areas include designing low-power chip for wireless applications, research on polar codes and related 5G standardization.
\end{IEEEbiography}

\begin{IEEEbiography}[{\includegraphics[width=1in,height=1.25in,clip,keepaspectratio]{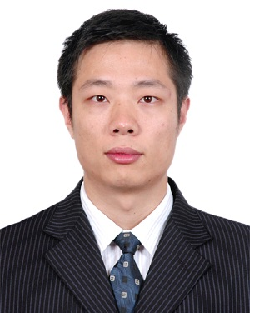}}]{Jun Wang}
is a senior research expert at Huawei Technologies Co., Ltd. His research areas include wireless communications, systems design and implementations.
\end{IEEEbiography}

\end{document}